Title

# Control of colloidal placement by modulated molecular orientation in nematic cells

Teaser:

Control of colloidal placement is achieved in nematic cells with photo-patterned spatially modulated molecular orientation.

**Authors**

Chenhui Peng[1], Taras Turiv[1], Yubing Guo[1], Sergij V. Shiyanovskii[1], Qi-Huo Wei[1], and Oleg D. Lavrentovich[1]*

Corresponding Author: *olavrent@kent.edu

**Affiliations**

[1]Liquid Crystal Institute and Chemical Physics Interdisciplinary Program, Kent State University, Kent, OH 44242, USA

**Abstract**

Colloids self-assemble into various organized superstructures determined by particle interactions. There is a tremendous progress in both the scientific understanding and applications of self-assemblies of single-type identical particles. Forming superstructures in which the colloidal particles occupy predesigned sites and remain in these sites despite thermal fluctuations represents a major challenge of the current state-of-the art. Here we propose a versatile approach to direct placement of colloids using nematic liquid crystals with spatially varying molecular orientation pre-imposed by substrate photoalignment. Colloidal particles in nematic environment are subject to the long-range elastic forces originating in the orientational order of the nematic. Gradients of the orientational order create an elastic energy landscape that drives the colloids into locations with preferred type of deformations. As an example, we demonstrate that colloidal spheres with perpendicular surface anchoring are driven into the regions of maximum splay, while spheres with tangential surface anchoring settle into the regions of bend. Elastic forces responsible for preferential placement are measured by exploring overdamped dynamics of the colloids. Control of colloidal self-assembly through patterned molecular orientation opens new opportunities for design of materials and devices in which particles should be placed in pre-designed locations.

One sentence summary:
Control of colloidal placement is achieved in nematic cells with photo-patterned spatially modulated molecular orientation.

# MAIN TEXT

## INTRODUCTION

Self-assembled colloidal structures are formed by small particles that interact in a predefined manner to create complex patterns (*1*). Self-assembly in isotropic solutions such as water already demonstrated a great potential to design materials for various scientific and technological purposes (*2*). Colloidal interactions and thus complexity of superstructures can be dramatically enriched by the use of liquid crystals (LCs) as a dispersion medium (*3*). Long-range orientational order of LCs leads to anisotropic colloidal interactions that control assembly into clusters (*4-13*), mediate interactions of particles with flat (*14*), curved (*15*), and templated interfaces (*16, 17*), cause trapping at topological defects (*18-23*). The spectrum of possible design pathways of colloidal self-assembly through LC dispersions is further expanded by sensitivity of the liquid crystalline order to external electromagnetic fields, temperature, or confinement.

Colloidal self-assembly in both isotropic and anisotropic media is currently well explored for the case when the particles are of the same type and occupy locations determined by interactions with other particles and bounding surfaces. The challenge is to develop robust approaches to place particles into predesigned locations. Current methods of heterogeneous design are based on lock-and-key colloidal interactions of non-spherical particles (*24*), programmable interactions of DNA molecules attached to the surface of particles (*25*), and using substrates with patterned magnetic (*26*) or hydrophobic properties (*27*). In this work, we propose a new approach to placement of colloidal particles that is based on LCs with predesigned molecular orientation, achieved through the substrate photo-patterning.

We use the simplest type of a LC, the so-called nematic, in which elongated molecules are oriented along a single direction called the director $\hat{\mathbf{n}}$, with the properties $\hat{\mathbf{n}} \equiv -\hat{\mathbf{n}}$ and $\hat{\mathbf{n}}^2 = 1$. The flat nematic cells bounded by two glass plates are designed in such a way that the director field experiences periodic one-dimensional distortions with alternating stripes of splay and bend. These deformations are pre-imposed by the substrate photo-alignment technique developed recently (*28, 29*). Spherical particles demonstrate a dramatic selectivity in spatial locations within the pre-imposed landscape of director deformations, depending on their surface anchoring properties, i.e., the preferred direction of the local $\hat{\mathbf{n}}$. Namely, spheres with tangential surface alignment of $\hat{\mathbf{n}}$ are attracted to the regions of bend, while spheres with perpendicular alignment are attracted to the regions of splay. Analysis of the colloidal dynamics allows us to determine the interaction potential of the particles with the underlying director landscape.

**RESULTS**

We explore a stripe pattern with periodic splay $\text{div}\hat{\mathbf{n}} \neq 0$ and bend $\hat{\mathbf{n}} \times \text{curl}\hat{\mathbf{n}} \neq 0$ modulated along the $y$-axis; the director is everywhere parallel to the bounding plates, thus there is no $z$-component, $n_z = 0$; the director field imposed by the surface photo-alignment layer is of the form

$$\hat{\mathbf{n}} = (n_x, n_y, n_z) = (\cos\alpha, \sin\alpha, 0), \tag{1}$$

where $\alpha = \pi y / l$, $l = 80\,\mu\text{m}$ is the period, Fig. 1. An experimental cell represents a thin (thickness $h = 20\,\mu\text{m}$) flat layer of a nematic sandwiched between two glass plates with identical photo-induced alignment patterns at the top ($z = h$) and the bottom ($z = 0$) surfaces. Splay and bend contribute to the Oseen-Frank free energy density of the distorted nematic,

$$f = \frac{1}{2}K_1(\text{div}\hat{\mathbf{n}})^2 + \frac{1}{2}K_2(\hat{\mathbf{n}}\cdot\text{curl}\hat{\mathbf{n}})^2 + \frac{1}{2}K_3(\hat{\mathbf{n}}\times\text{curl}\hat{\mathbf{n}})^2, \tag{2}$$

with the elastic constants of splay $K_1$, twist $K_2$, and bend $K_3$, respectively. For the director patterns Eq. (1) used in this work, twist (as well as saddle-splay) deformations vanish, since the director does not change along the $z$-axis. Figure 1B shows the spatial distribution of the dimensionless splay energy density $(l/\pi)^2 (\text{div}\hat{\mathbf{n}})^2 = \cos^2(\pi y/l)$ corresponding to Eq. (1). As detailed by the experiments below, the alternating regions of splay and bend attract colloidal spheres with different surface anchoring, namely, spheres with normal anchoring are attracted to the splay regions, Fig. 1, and spheres with tangential anchoring are attracted to the bend regions, as discussed later.

**Spheres with normal anchoring**. Polystyrene spheres of radius $R = 2.5\,\mu\text{m}$ were covered with a thin layer of a surfactant to produce perpendicular (also called homeotropic, or normal) alignment of the director at their surface. When such a sphere is placed in a nematic with a uniform director field $\hat{\mathbf{n}}_0 = (1, 0, 0)$, the radial-like director around the sphere is topologically compensated by a point defect, the so-called hyperbolic hedgehog located either on the right or left hand side of the sphere (*3, 30*). The resulting director structure is in the form of a dipole, Fig. 1A. We direct the unit vector $\hat{\mathbf{p}}$ of the dipole from the hedgehog towards the sphere. Note that once the direction of the dipole is randomly selected to be either parallel or antiparallel to the $x$-axis of the uniform cell with $\hat{\mathbf{n}}_0 = (1, 0, 0)$, it cannot be reversed into an opposite orientation, as the energy barrier for such a realignment, on the order of $KR \sim (10^{-16} - 10^{-17})\,\text{J}$ is much larger than the thermal energy $k_B T = 4.2 \times 10^{-21}\,\text{J}$, with the typical value of the Frank elastic constant $K$ being a few tens of pN. In principle, in very shallow cells, of a thickness comparable to the colloid diameter, the dipolar structure can transform into a Saturn-ring configuration (*31*); however, in all our experiments the dipolar structure remained stable (*32, 33*). This implies that the homeotropic anchoring is strong, with the anchoring coefficient being larger than $K/R \sim 10^{-5}\,\text{J/m}^2$.

Individual spheres with normal anchoring placed in the cell with a periodic splay and bend migrate towards the regions with the maximum splay, $y = 0, \pm l, \pm 2l, ...$, Fig. 1B. When the concentration of spheres is high enough, they form linear chains along the $x$-axis, located at the same ordinate of the maximum splay, $y = 0, \pm l, \pm 2l, ...$ Fig. 1C, Fig. S1.

It is important to note a polar character of colloidal assemblies in the splay regions. In a uniform nematic, "left" and "right" orientations of $\hat{\mathbf{p}}$ are equally probable. In the described splay-bend pattern, which lacks the left-right mirror symmetry, all the dipoles align themselves along the same direction, forming a ferroelectric type of order. For the chosen geometry of the splay-bend pattern, Eq. (1) and Fig. 1B, the dipoles $\hat{\mathbf{p}}$ of the spheres trapped at the splay locations $y = 0, \pm l, \pm 2l, ...$ are always parallel to the $x$-axis. If we use an optical tweezers and place a sphere in any location within the pattern specified by Eq. (1), then the sphere orients its dipole $\hat{\mathbf{p}}$ parallel or antiparallel to the local director. If the sphere is released from the optical tweezers, it moves towards the regions $y = 0, \pm l, \pm 2l, ...$, while simultaneously rotating the dipole in accordance with the rotating local director orientation, Fig. 1B. In particular, the tweezers can place a sphere in the splay region in such a way that the dipole $\hat{\mathbf{p}}$ is antiparallel to the $x$-axis. This configuration is unstable, since $\hat{\mathbf{p}}$ prefers to be parallel to the $x$-axis at equilibrium. The sphere with the "wrong" orientation of $\hat{\mathbf{p}}$ moves across the stripe pattern, rotating $\hat{\mathbf{p}}$ in accordance with the local director, until it reaches the neighboring splay region with $\hat{\mathbf{p}}$ being parallel to the $x$-axis, i.e. completing a 180 degrees rotation through a shift by $|\Delta y| = l$, Fig. 1D and 1E. Note that simple reorientation of $\hat{\mathbf{p}}$ by 180 degrees with the sphere keeping its $y$-coordinate unchanged, $|\Delta y| = 0$, is never observed in our experiments; the reason is that such a process is associated with a huge energy barrier $KR >> k_B T$, similar to the case of the uniform cell discussed above.

At moderate concentrations, colloidal particles with homeotropic anchoring gather into linear chains, Fig.1C. Note that all the chains have the same direction of the individual dipole moments $\hat{\mathbf{p}}$, being separated by the same distance $l$ along the $y$-axis. By increasing the concentration of colloids, the length of the chains can be increased, but since we are interested primarily in the individual behavior, we limit the study by relatively low concentrations, in which the number of particles in each chain is less than about 10.

The interaction potential between a colloidal sphere with a dipolar director and a distorted director has been derived by Pergamenshchik (*34*). We follow Ref. (*34*), but use a modified expression that conforms to our definition of the dipole; for a sphere of radius $R \ll l$, the interaction potential writes

$$U_d = -8\pi\beta K R^2 \left(\hat{\mathbf{p}} \cdot \hat{\mathbf{n}}\right) \mathrm{div}\,\hat{\mathbf{n}}, \tag{3}$$

where $\beta$ is the amplitude of dipole director distortions that depends on the strength of surface anchoring, anisotropy of elastic constants, accuracy of the dipolar approximation, etc.; $K$ is the effective elastic constant in the so-called one-constant approximation. In the analysis below we treat $\beta$ as an unknown numerical coefficient, which we determine from the fitting of experimental data. In Eq. (3), we introduce the scalar product $\left(\hat{\mathbf{p}} \cdot \hat{\mathbf{n}}\right)$ to account for the polar character of coupling between the splay $\mathrm{div}\,\hat{\mathbf{n}}$ and $\hat{\mathbf{p}}$ and also to preserve the nematic symmetry $\hat{\mathbf{n}} \equiv -\hat{\mathbf{n}}$. For the director field in Eq. (1), $\hat{\mathbf{n}} = \left[\cos(\pi y/l), \sin(\pi y/l), 0\right]$, the interaction potential for a given sphere that keeps $\left(\hat{\mathbf{p}} \cdot \hat{\mathbf{n}}\right) = 1$, namely, $U_d = -8\pi^2 \beta K \left(R^2/l\right)\cos(\pi y/l)$, reaches its minimum $U_{d,\min} = -8\pi^2 \beta K \left(R^2/l\right)$ at the locations $y = 0, \pm 2l, \ldots$, and maximum $U_{d,\max} = 8\pi^2 \beta K \left(R^2/l\right)$ at the locations $y = \pm l, \pm 3l, \ldots$. In other words, the normally anchored spheres are expected to gather in the regions with a maximum splay, being oriented in such a way that the dipole $\hat{\mathbf{p}}$ fits the polarity of splay $\hat{\mathbf{n}} \operatorname{div} \hat{\mathbf{n}}$.

The potential Eq. (3) also explains the behavior of a sphere placed by the optical tweezers in a splay region with a "wrong" polarity. Such a sphere moves along the potential energy landscape from the maximum energy at $y = l$ towards the minimum at $y = 0$, Fig. 1D and 1E. The dynamics of the sphere from $y = l$ towards $y = 0$ can be used to extract the information about the energetics of elastic coupling between the colloids and the surrounding director distortions, as detailed below.

The particle's dipole preserves its orientation parallel to the local director during migration. Then the elastic force driving the sphere to the minimum energy location writes

$$\mathbf{F} = -\left(\partial U_d / \partial y\right)\hat{\mathbf{e}}_y = -8\pi^3 \beta K \left(\frac{R}{l}\right)^2 \sin\left(\frac{\pi y}{l}\right)\hat{\mathbf{e}}_y, \tag{4}$$

where $\hat{\mathbf{e}}_y$ is the unit vector along the $y$-axis. Since the Reynolds number in our experiments is small and inertia contributions can be neglected, the driving force is balanced by the viscous drag force $\mathbf{F}_d = -6\pi R \ddot{\eta} \mathbf{v}$, where $\ddot{\eta}$ is the diagonal viscosity tensor with the components $\eta_\parallel$ and $\eta_\perp$, $\mathbf{v}$ is the particle velocity vector with the components $v_\parallel$ and $v_\perp$; the subscripts $\parallel$ and $\perp$ refer to the components parallel and perpendicular to the director, respectively. The drag force can be rewritten in terms of the components of the viscosity tensor, $\mathbf{F}_d / 6\pi R = \eta_\parallel v_\parallel \left(\hat{\mathbf{e}}_x \cos\alpha + \hat{\mathbf{e}}_y \sin\alpha\right) + \eta_\perp v_\perp \left(-\hat{\mathbf{e}}_x \sin\alpha + \hat{\mathbf{e}}_y \cos\alpha\right)$. Balancing the driving and drag forces in the laboratory coordinate system, we obtain the equations for the velocity components $v_x = v_\parallel \cos\frac{\pi y}{l} - v_\perp \sin\frac{\pi y}{l}$ and $v_y = v_\parallel \sin\left(\frac{\pi y}{l}\right) + v_\perp \cos\left(\frac{\pi y}{l}\right)$ in the following form:

$$\frac{\partial x}{\partial t} = -\frac{4\pi^2}{3} \frac{\eta_\perp - \eta_\parallel}{\eta_\parallel \eta_\perp} \beta \frac{KR}{l^2} \sin^2 \frac{\pi y}{l} \cos \frac{\pi y}{l}, \tag{5}$$

$$\frac{\partial y}{\partial t} = -\frac{4\pi^2}{3} \beta \frac{KR}{l^2} \frac{\eta_\perp \sin^2 \frac{\pi y}{l} + \eta_\parallel \cos^2 \frac{\pi y}{l}}{\eta_\perp \eta_\parallel} \sin \frac{\pi y}{l}. \tag{6}$$

Solutions of Eq. (5) and Eq. (6) establish how the sphere's coordinates $x$ and $y$ vary with time. These theoretical dependencies are used to fit the experimental trajectory in Fig. 1F and 1G in order to verify the model and to deduce the unknown parameter $\beta$.

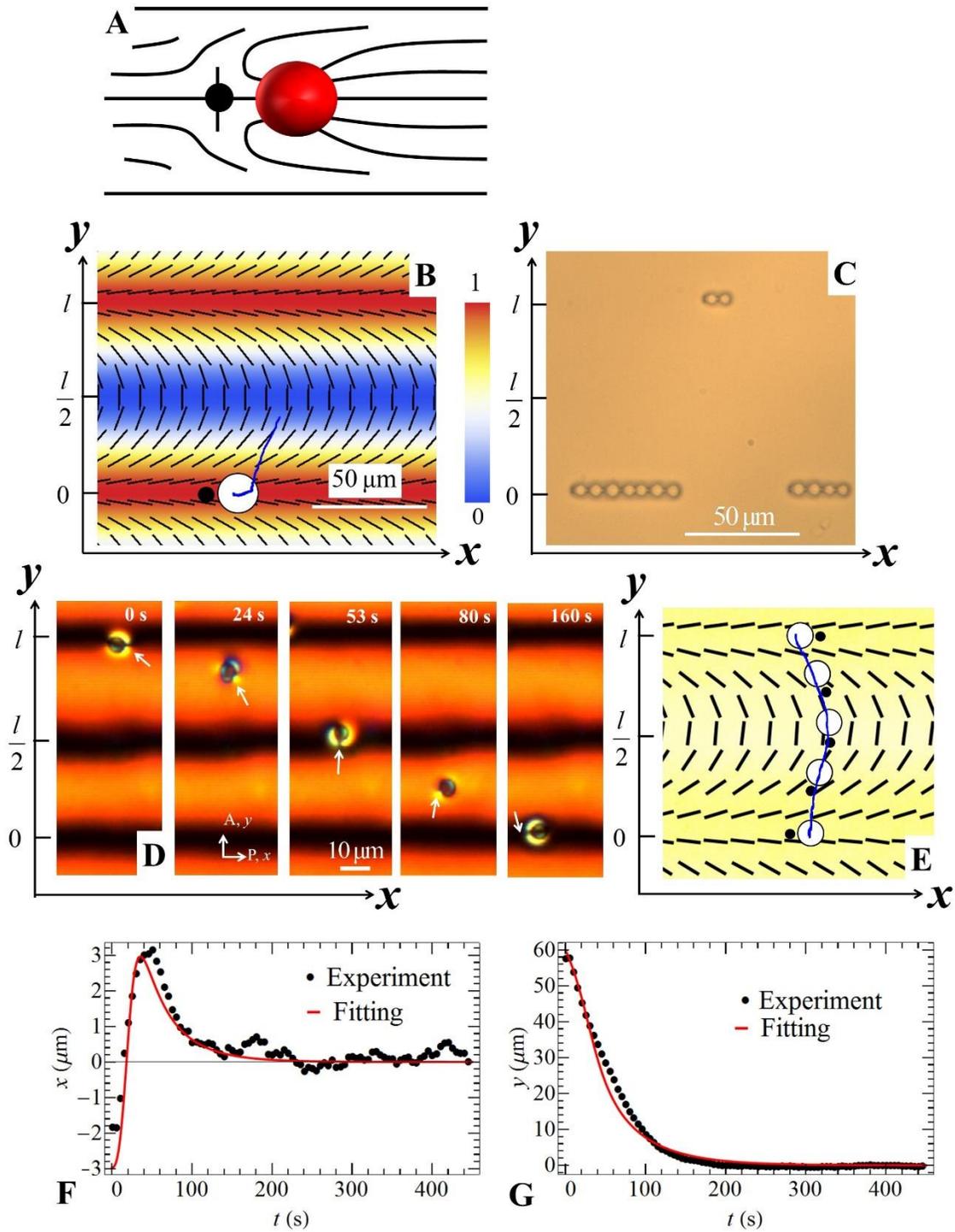

**Fig.1. Elasticity-directed placement and assembly of normally anchored polystyrene spheres in regions with splay deformations of the patterned director.** (**A**) A sphere with perpendicular surface anchoring placed in a uniform nematic causes an appearance of a hyperbolic hedgehog that

can be located on either the left or right hand side; the resulting director deformations are of a dipolar type ; (**B**) Periodic splay-bend stripe pattern; the normalized splay energy density is labeled by colors according to the scale on the right hand side. A sphere placed by optical tweezers in the bend region and released there, migrates towards the splay region. The dark blue curve is the typical experimental trajectory of the sphere's center. (**C**) Self-assembly of spheres into linear chains in the regions of maximum splay with $p > 0$ (bright field microscope, unpolarized light, nematic liquid crystal pentylcyanobiphenyl (5CB)); (**D**) Polarizing microscopy images of a sphere moving from one splay region to the next splay region, with concomitant reorientation of the structural dipole from $p < 0$ to $p > 0$; the white arrows point towards the hyperbolic hedgehog. (**E**) Typical experimental trajectory of the sphere within the director configuration mapped in the PolScope mode of observation of the experimental cell. (**F**) Experimentally measured $x(t)$ dependence for a sphere moving from $y = l$ to $y = 0$ and its theoretical fit by Eq. (5). (**G**) Experimentally measured $y(t)$ dependence for a sphere moving from $y = l$ to $y = 0$ and its theoretical fit by Eq. (6). All spheres are of a radius $R = 2.5\,\mu m$; the period of patterned director is $l = 80\,\mu m$. The nematic material corresponding to parts (D-G) is MLC6815.

========================================

Fitting of the experimental dynamics of the colloids requires a knowledge of the two components of the viscosity tensor, $\eta_\parallel$ and $\eta_\perp$. These two were determined experimentally for the same liquid crystal and normally anchored spheres by measuring the mean square displacements (MSD) as a function of time for the Brownian diffusion parallel and perpendicular to the director in uniformly aligned cells, $\hat{\mathbf{n}}_0 = (1,0,0)$, Fig. 2. The slopes of the MSD vs. time dependencies yield the diffusion coefficient $D_\parallel = (7.6 \pm 0.2) \times 10^{-15}\,m^2/s$ for the direction parallel to $\hat{\mathbf{n}}_0$ and $D_\perp = (4.1 \pm 0.2) \times 10^{-15}\,m^2/s$ for the perpendicular direction. These values yield the corresponding components of the viscosity tensor, $\eta_\parallel = (12\pm 1)\,\text{mPa}\cdot\text{s}$ and $\eta_\perp = (22\pm 1)\,\text{mPa}\cdot\text{s}$, which follow from the Einstein-Stokes formula $\eta = k_B T / (6\pi R D)$.

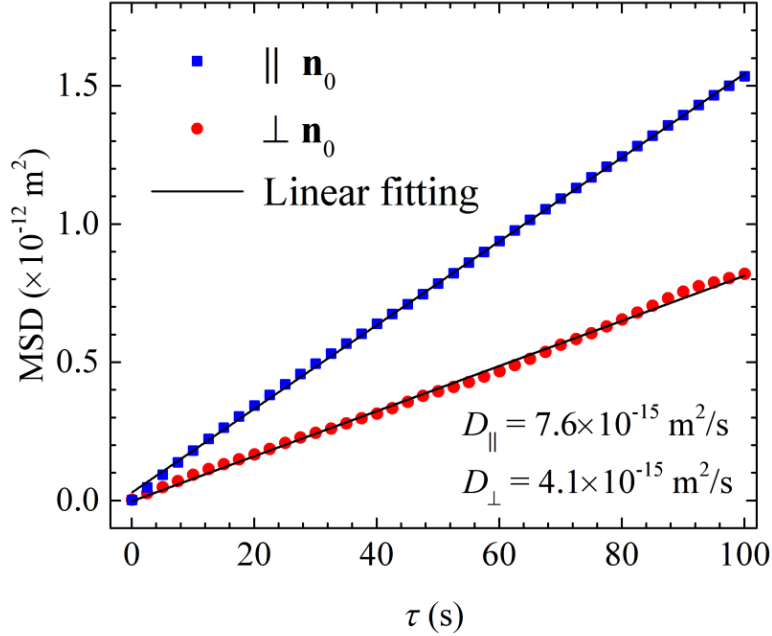

**Fig. 2. Mean squred displacement (MSD) of a dipolar colloid in uniformly aligned LC cell.** Homeotropically anchored sphere of radius $2.5\,\mu\text{m}$ in a nematic MLC6815 cell with uniform alignment $\hat{\mathbf{n}}_0$, which is achieved by unidirectional rubbing of the aligning layer of PI2555 deposited on the bounding plates.

========================================

As clearly seen from Fig. 1F and 1G, the experimental data are closely matched by the expected theoretical behavior, despite the fact that the model uses the so-called one-constant approximation. The only fitting parameter is $\beta = 0.20 \pm 0.02$. Note that with this value and with $K = 10^{-11}\,\text{N}$, the depth of the potential minimum, $U_{d,\max} - U_{d,\min} = 16\pi^2 \beta K (R^2/l)$ is $2.2\times 10^{-17}\,\text{J}$, i.e., about four orders of magnitude higher than the thermal energy $k_B T = 4.2\times 10^{-21}\,\text{J}$. In other words, the elastic trapping is much stronger than the randomizing forces associated with thermal fluctuations. One can also estimate the stiffness of the potential trap that keeps the normally anchored spheres in the splay regions of a "correct" polarity. For $y \to 0$, the stiffness of the elastic trap $k = \frac{\partial^2 U_d}{\partial y^2} = 8\pi^4 \beta \frac{R^2}{l^3} K$ is about $1.9 \times 10^{-8}$ N/m.

**Spheres with tangential anchoring.** Polystyrene spheres without surfactant coatings produce spontaneous tangential (parallel to the interface and degenerate) anchoring. The director deformations around these spheres in a uniform nematic are of quadrupolar symmetry, with two point defects-boojums at the poles, Fig. 3A. Placing these spheres in the splay-bend stripe pattern, Eq. (1), produces behavior qualitatively different from the normally anchored spheres, Fig.3. Fig. 3B maps the dimensionless bend energy $(l/\pi)^2 (\hat{\mathbf{n}} \times \mathrm{curl}\hat{\mathbf{n}})^2 = \sin^2(\pi y/l)$ stored in the stripe pattern. The tangentially anchored spheres migrate towards the regions with the maximum bend, i.e. locations with $y = \pm l/2, \pm 3l/2,...$, Fig.3B, 3C. As the concentration of spheres increases, they assemble into chains with tilted arrangements of the neighboring spheres, similar to those described for uniform nematic cells in Ref. (*35*). Figure 3C shows only the short chains as we are primarily interested in the individual behavior of colloids in the patterned landscape. Longer chains form in more concentrated dispersions and can extend over a length of 50-70 μm across the region with bend distortions; they are disrupted when their ends become close to the regions with splay deformations.

The potential of interaction between a quadrupolar particle and the distorted director field reads, following Pergamenshchik (*34*), as

$$U_Q = -4\pi K \chi R^3 (\hat{\mathbf{n}} \cdot \nabla) \mathrm{div}\hat{\mathbf{n}}, \tag{7}$$

where $\chi$ is the dimensionless coefficient of the quadrupolar moment, the value of which is unknown and depends on the surface anchoring. For the patterned director field, Eq. (1), this potential reduces to $U_Q = 4\pi^3 \chi K \dfrac{R^3}{l^2} \sin^2 \dfrac{\pi y}{l}$, thus predicting that the spheres with $\chi < 0$ will minimize the potential by migrating towards the locations with bend, $y = \pm l/2, \pm 3l/2,...$ as observed in the experiments, Fig. 3C. The corresponding force driving the spheres to the equilibrium locations is

$$F_Q = -\left(\partial U_Q / \partial y\right) = -4\pi^4 \chi K \left(\dfrac{R}{l}\right)^3 \sin \dfrac{2\pi y}{l}. \tag{8}$$

If the sphere is placed by optical tweezers in the splay region $y = 0, \pm l,...$, a fluctuative displacement from that location causes the sphere to move towards the bend stripe region at $y = \pm l/2, \pm 3l/2...,$, as shown by a trajectory in Fig. 3B. By balancing the driving and drag forces, one finds the $x$-component and $y$-component of the velocity, similarly to the case of normally anchored spheres:

$$\dfrac{\partial x}{\partial t} = -\dfrac{\pi^3}{3} \dfrac{\eta_\perp - \eta_\parallel}{\eta_\parallel \eta_\perp} \chi \dfrac{KR^2}{l^3} \sin^2 \dfrac{2\pi y}{l}, \tag{9}$$

$$\dfrac{\partial y}{\partial t} = -\dfrac{2\pi^3}{3} \chi \dfrac{KR^2}{l^3} \dfrac{\eta_\perp \sin^2 \dfrac{\pi y}{l} + \eta_\parallel \cos^2 \dfrac{\pi y}{l}}{\eta_\parallel \eta_\perp} \sin \dfrac{2\pi y}{l}. \tag{10}$$

Proceeding as above, by solving Eq. (9) and (10) and finding the theoretical trajectory of the particle, we fit the experimental dependencies $x(t)$ and $y(t)$, Fig. 3D and 3E. The model describes the experimental behavior well, and yields the value of quadrupolar parameter $\chi = -0.95 \pm 0.10$. With this value, the depth of the trapping potential $U_{Q,\max} - U_{Q,\min} = 4\pi^3 |\chi| \dfrac{KR^3}{l^2}$ can be estimated as $2.9 \times 10^{-18} \text{J}$, which is much larger than the thermal energy. Finally, we can estimate the stiffness $k = \left.\dfrac{\partial^2 U_Q}{\partial y^2}\right|_{y \to l/2} = 8\pi^5 K |\chi| \dfrac{R^3}{l^4}$ of the potential, which is $9 \times 10^{-9}$ N/m, i.e., about two times weaker than the stiffness of the splay trap acting on the normally anchored spheres.

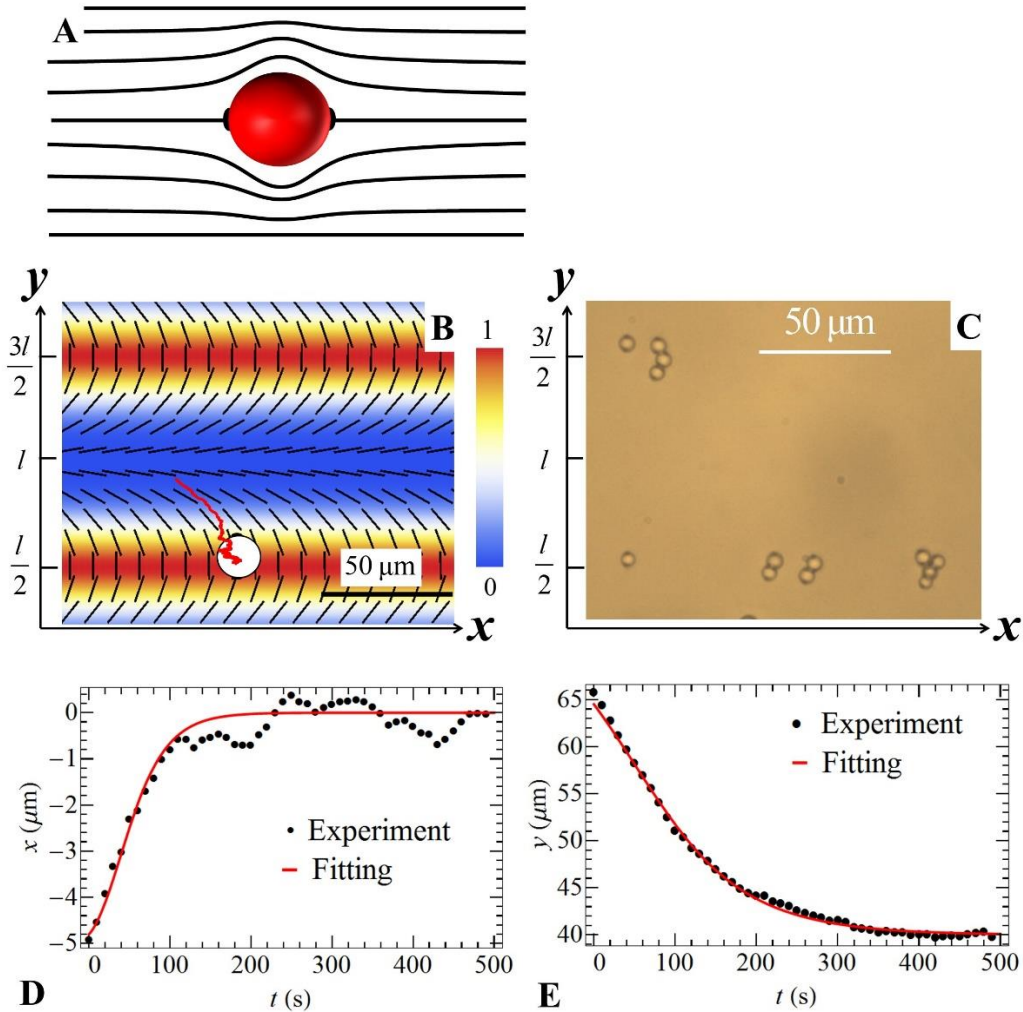

**Fig. 3. Elasticity-directed placement and assembly of tangentially anchored polystyrene spheres in regions with bend deformations of the patterned director.** (**A**) Director deformations around a tangentially anchored sphere in a uniform nematic are of quadrupolar symmetry, with two point defects-boojums at the poles; (**B**) Periodic splay-bend stripe pattern; the normalized bend energy density is labeled by colors according to the scale on the right hand side. A sphere placed by optical tweezers in the splay region and released there, migrates towards the bend region. The dark red curve is the typical experimental trajectory of the sphere's center. (**C**) Self-assembly of spheres into chains in the regions of maximum bend (bright field microscope,

unpolarized light). (**D**) Experimentally measured $x(t)$ dependence for a sphere moving from $y = l$ to $y = l/2$ and its theoretical fit by Eq. (9). (**E**) Experimentally measured $y(t)$ of a sphere moving from $y = l$ to $y = l/2$ and its theoretical fit by Eq. (10). All spheres are of a radius $R = 2.5\,\mu\text{m}$; periodicity of the director pattern is $l = 80\,\mu\text{m}$. Liquid crystal used is 5CB in (C) and MLC6815 in (D, E).

**Preferential placement of spheres with different surface anchoring.** So far, we discussed dispersions of single type particles. When the cell is filled with a mixture of two types of particles, the surface-patterned director landscape leads to their preferential placement in different regions of the cell, Fig. 4. Namely, the normally anchored spheres migrate towards the splay, $y = 0$, and the tangentially anchored spheres migrate towards the bend, $y = \pm l/2$, Fig. 4. Depending on the overall concentration of the spheres, they can form larger or smaller aggregates. The normally anchored spheres produce linear chains parallel to the direction of stripes, while the tangentially anchored spheres form chains that are predominantly perpendicular to this direction, which is an expected behavior, as evidenced by prior studies of aggregation of both types of spheres in uniformly aligned nematic cells (*3*), Fig. 4.

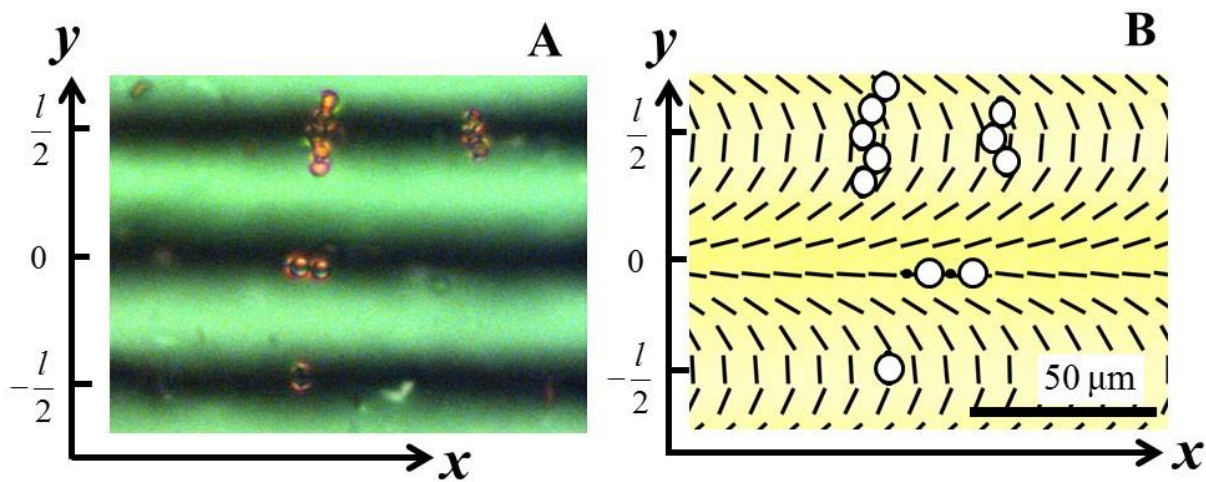

**Fig. 4. Preferential placement of colloids with different surface properties in a cell with periodically patterned splay-bend director field.** (**A**) Polarizing microscope texture of polystyrene colloids with homeotropic and tangential anchoring located in the alternating bands of splay and bend, respectively. (**B**) Scheme of preferential placement with the experimentally determined director pattern (PolScope observations). The homeotropically anchored spheres reside in the region of splay, $y = 0$, while tangentially anchored spheres localize in the regions of bend, $y = \pm l/2$. Note different orientation of chains formed by homeotropic spheres (along the $x$-axis) and by the tangential spheres (perpendicular to the $x$-axis). The colloids are placed in a LC mixture with zero dielectric anisotropy. P and A show the directions of polarizer and analyzer.

===================

## DISCUSSION

In this work, we demonstrate a new approach to control colloidal placement, by using the long-range elastic forces in spatially modulated nematic LC. The approach is based on the pre-designed molecular orientation pattern at the bounding plates of the cell. The colloidal spheres interact with the underlying director distortions and migrate into the regions where the elastic energy of interactions is minimized. For spheres with normal anchoring these regions correspond to the splay deformations; tangentially anchored spheres, in contrast, are localized in the regions of bend.

The periodic splay-bend director pattern imposes strong limitations on the collective behavior of the colloids. For example, the spheres with normal anchoring form a ferroelectric type of macroscopic ordering, in which the structural dipole moment $\hat{\mathbf{p}}$ of each and every particle points into the same direction. The system is very deterministic, as we do not observe any "mistakes" made by the colloids, for example, wrong orientation of $\hat{\mathbf{p}}$ or wrong placement of the sphere with the tangential anchoring (which would be in the regions of splay). Note here that we explored four different nematic materials and all of them showed similar behavior, which underscores a general and robust character of the colloidal interactions mediated by surface anchoring and elasticity of the pre-imposed director patterns.

Because the period $l$ of the splay-bend background is much bigger than the particle radius $R$, the interaction potential is described adequately by the lowest multipole deformation around the particles that grows as $R^2/l$ for the spheres with locally dipolar director configuration around them, and as $R^3/l^2$ for quadrupolar spheres with tangential anchoring. Thermal fluctuations do not affect the particle dynamics in our experiments. For the given strength of director gradients, defined by the period $l = 80\ \mu m$, the smallest radius of the spheres that can be trapped at the desired location in space, can be estimated by comparing the depth of the trapping potentials to the thermal energy. For the dipolar particles, the smallest radius is estimated from the condition $U_{d,\max} - U_{d,\min} = k_B T$, to be 34 nm. The last estimate presumes that the surface anchoring strength $W$ is sufficiently strong to maintain the dipolar configuration, namely, $K/W \sim 10$ nm or so. For the quadrupolar particles, placed in a pattern with $l = 80\ \mu m$, the minimum size is about 220 nm. The feature of a relatively large size limit follows from the additional factor $R/l$ in the quadrupolar potential as compared to the dipolar potential. By making the pattern with a smaller period $l$, one can decrease the minimum size of the controllable particles below the estimates above. To describe the case of smaller $l$ accurately, however, one needs to account for (i) the finite size of the particles and (ii) finite surface anchoring strength at the surface of the spheres and at the bounding plates.

The presented experimental data on dynamics and placement of the colloidal particles are well described by the elastic model based on the Frank-Oseen functional and the dipolar and quadrupolar type of director deformations around the particles. One of the reason why the experiment and the theory match well is the relatively large size of the colloids, $5\,\mu m$. For submicron particles and for high concentration of particles, the description might require a more elaborate approach that would, for example, account for the finite anchoring strength. Furthermore, the theoretical description might be supplemented by the consideration of thermal fluctuations. In the uniformly aligned liquid crystal environment, displacement of particles caused by Brownian motion are both anisotropic(*36*) and, at time scales on the order of the relaxation time of the director, also anomalous (*37*). It would be of interest to explore how the pre-patterned director deformations might influence the regimes of colloidal Brownian motion.

**CONCLUSIONS**

To conclude, the photo-patterning surface alignment offers a versatile method to control colloidal assemblies by directing particles into predesigned locations and trapping them in these locations with forces that exceed those of entropic nature. Besides the practical utility, the study clearly demonstrates and quantifies experimentally the mechanisms of selective interactions between the colloids and the surrounding landscape of a distorted director. In this work, we demonstrated the principle using spheres and a relatively simple periodic director pattern with alternating splay and bend. It is obvious, however, that the similar approach can be extended to particles of different shape or size and to more complex 2D and even 3D patterns of the director field. We also expect that the approach is not limited by just normal and tangential anchoring at the surface of the colloidal particles; tilted anchoring is also expected to show selectivity in equilibrium placement within a patterned director field. Colloidal placement in the photo-patterned templates can be realized at practically any spatial scale without mechanical stimuli (e.g., shearing, pressure gradients) or topographically modified surfaces (*38*). The described ability to deterministically pre-design locations of colloidal particles of different types in self-assembled patterns can be of importance for the development of microfluidic, electro-optical and sensing devices. One of the applications might be sorting of particles with different surface properties and chemical composition, which we currently explore.

**Materials and Methods**

**Materials.** We use the nematic 5CB, MLC6815 and a LC mixture with zero dielectric anisotropy ($|\Delta\varepsilon| \leq 10^{-3}$) formulated by MLC7026-000 and E7 (in weight proportion 89.1:10.9) (all materials are purchased from *EM Industries*). The LCs are doped with 0.01wt% of polystyrene colloids of radius $2.5\,\mu m$ (*Duke Scientific*). Untreated colloids yield tangential anchoring. The colloids treated with octadecyl-dimethyl-(3-trimethoxysilylpropyl) ammonium chloride (DMOAP) produce perpendicular director alignment and dipolar structures with a hyperbolic hedgehog on one side of the sphere. The colloidal dispersion in the LC is injected into the photo-patterned cell with thickness $h = 20$ μm at room temperature $22^{\circ}C$.

**Optical microscopy.** We used Nikon TE2000 inverted microscope with both immersion (100x Fluor, N.A.=1.3) and dry (60x MPlan ELWD, N.A.= 0.7) objectives. Immersion objective was used with the optical tweezers (green laser Verdi V6, 532 nm, 10 mW) mounted with the microscope to move the particle off the equilibrium state. Polarizing microscopic images are captured by using camera Energent HR20000 and the dynamics images are taken by the monochromatic camera MotionBlitz Eomini2 with different frame rates (1-20 fps).

**PolScope microscopy.** The director fields produced by photo-patterning are established with the help of a microscope (Nikon E600) equipped with Cambridge Research Abrio LC-PolScope package. The PolScope uses a monochromatic illumination at 546 nm and maps optical retardance and orientation of the optical axis (*39*).

**Acknowledgments**

We thank Juan de Pablo, Oleh Tovkach, Rui Zhang, and Ye Zhou for useful discussions. **Funding:** This work was supported by National Science Foundation grants DMR-1507637, DMR-1121288 and CMMI-1436565. **Author contributions:** Conceived and designed the experiments: O. D. L., C. P., and Q.-H. W. Performed the experiments: C. P., T. T. and Y. G. Analyzed the data: C. P., T. T., S. S. and O. D. L. All authors participated in discussing and writing the manuscript.

# Figures

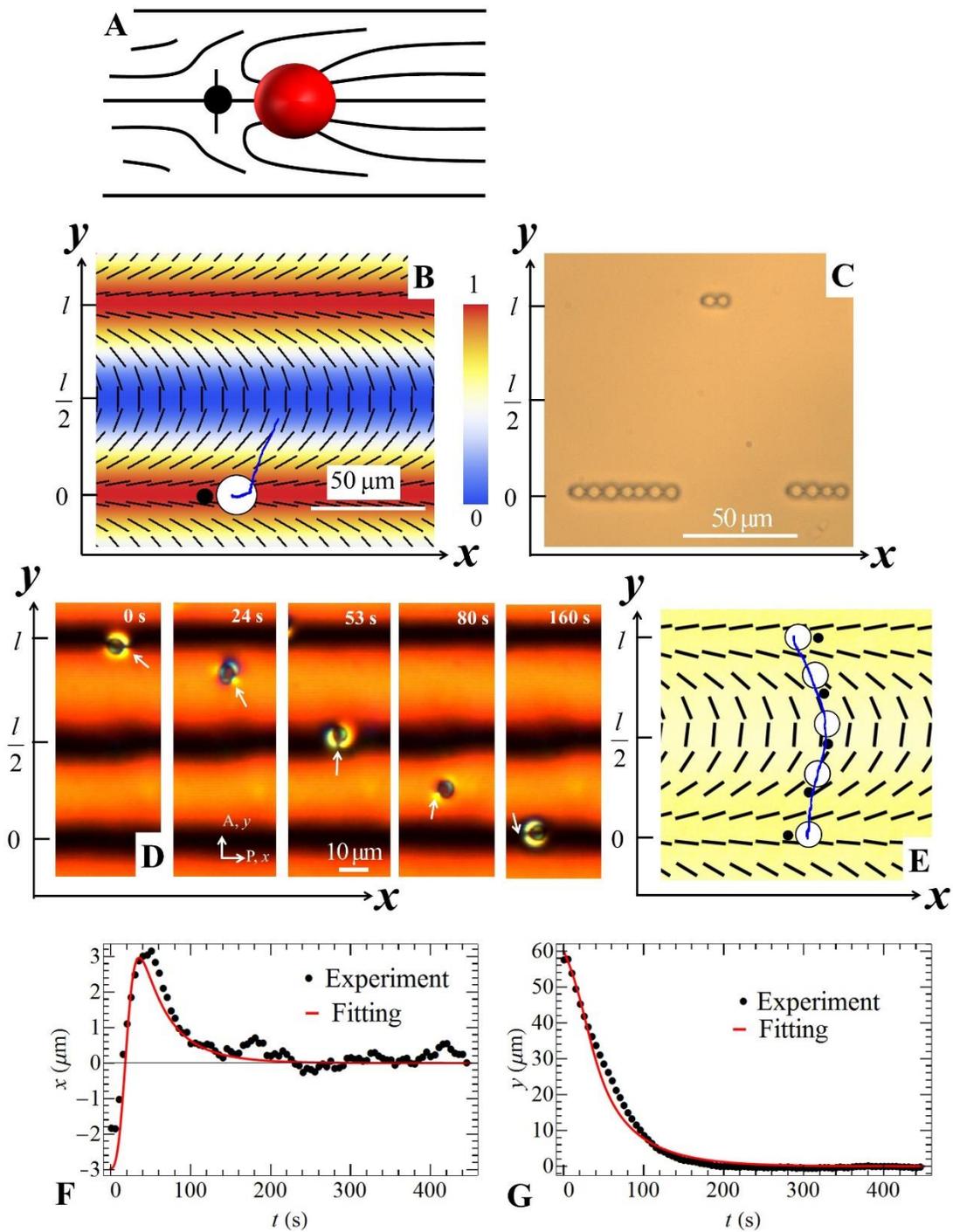

**Fig.1. Elasticity-directed placement and assembly of normally anchored polystyrene spheres in regions with splay deformations of the patterned director.** (**A**) A sphere with perpendicular surface anchoring placed in a uniform nematic causes an appearance of a hyperbolic hedgehog that

can be located on either the left or right hand side; the resulting director deformations are of a dipolar type ; (**B**) Periodic splay-bend stripe pattern; the normalized splay energy density is labeled by colors according to the scale on the right hand side. A sphere placed by optical tweezers in the bend region and released there, migrates towards the splay region. The dark blue curve is the typical experimental trajectory of the sphere's center. (**C**) Self-assembly of spheres into linear chains in the regions of maximum splay with $p > 0$ (bright field microscope, unpolarized light, nematic liquid crystal pentylcyanobiphenyl (5CB)); (**D**) Polarizing microscopy images of a sphere moving from one splay region to the next splay region, with concomitant reorientation of the structural dipole from $p < 0$ to $p > 0$; the white arrows point towards the hyperbolic hedgehog. (**E**) Typical experimental trajectory of the sphere within the director configuration mapped in the PolScope mode of observation of the experimental cell. (**F**) Experimentally measured $x(t)$ dependence for a sphere moving from $y = l$ to $y = 0$ and its theoretical fit by Eq. (5). (**G**) Experimentally measured $y(t)$ dependence for a sphere moving from $y = l$ to $y = 0$ and its theoretical fit by Eq. (6). All spheres are of a radius $R = 2.5\,\mu m$; the period of patterned director is $l = 80\,\mu m$. The nematic material corresponding to parts (D-G) is MLC6815.

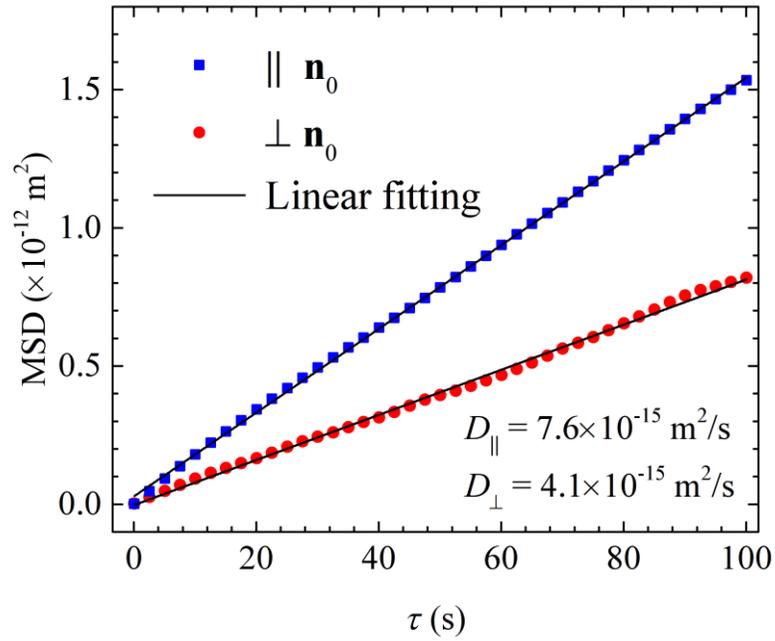

**Fig. 2. Mean squred displacement (MSD) of a dipolar colloid in uniformly aligned LC cell.** Homeotropically anchored sphere of radius 2.5 μm in a nematic MLC6815 cell with uniform alignment $\hat{\mathbf{n}}_0$, which is achieved by unidirectional rubbing of the aligning layer of PI2555 deposited on the bounding plates.

.

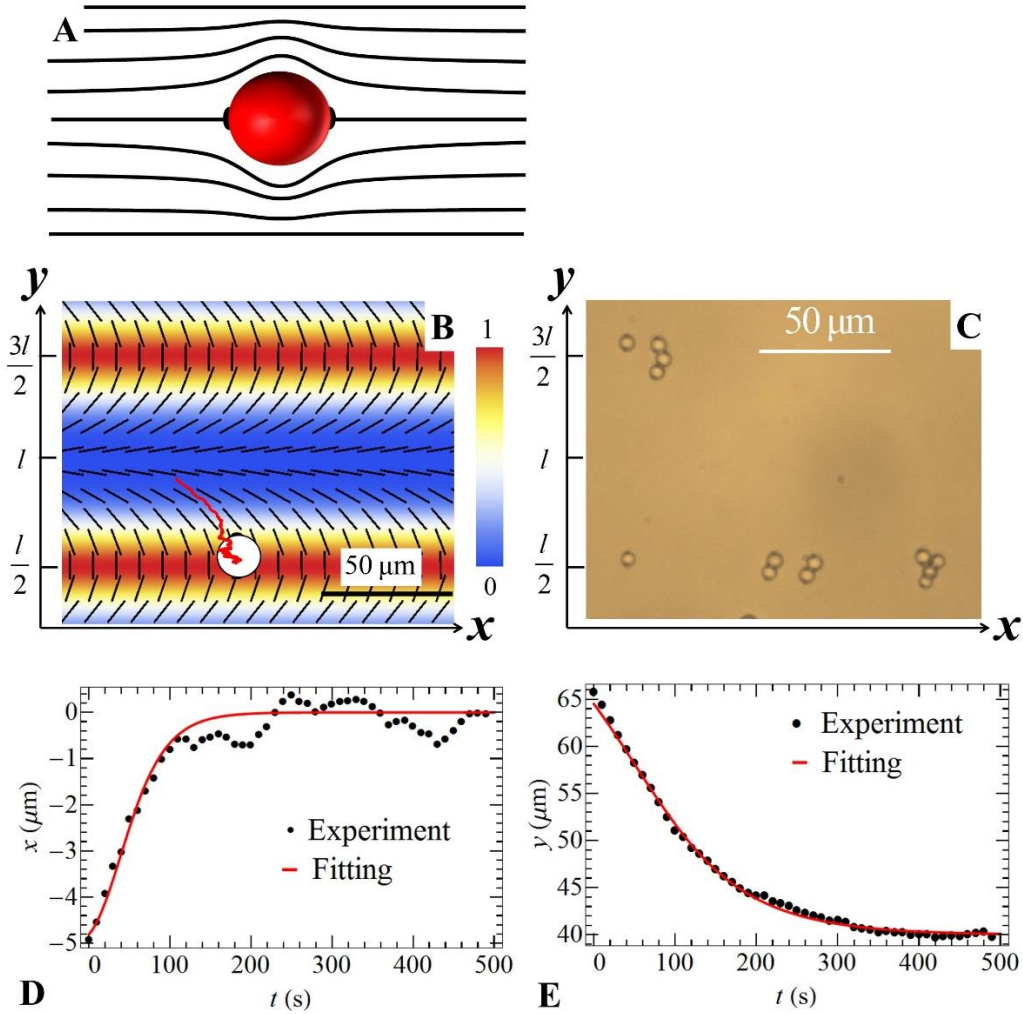

**Fig. 3. Elasticity-directed placement and assembly of tangentially anchored polystyrene spheres in regions with bend deformations of the patterned director.** (**A**) Director deformations around a tangentially anchored sphere in a uniform nematic are of quadrupolar symmetry, with two point defects-boojums at the poles; (**B**) Periodic splay-bend stripe pattern; the normalized bend energy density is labeled by colors according to the scale on the right hand side. A sphere placed by optical tweezers in the splay region and released there, migrates towards the bend region. The dark red curve is the typical experimental trajectory of the sphere's center. (**C**) Self-assembly of spheres into chains in the regions of maximum bend (bright field microscope, unpolarized light). (**D**) Experimentally measured $x(t)$ dependence for a sphere moving from $y = l$ to $y = l/2$ and its theoretical fit by Eq. (9). (**E**) Experimentally measured $y(t)$ of a sphere moving from $y = l$ to $y = l/2$ and its theoretical fit by Eq. (10). All spheres are of a radius

$R = 2.5\,\mu m$; periodicity of the director pattern is $l = 80\,\mu m$. Liquid crystal used is 5CB in (C) and MLC6815 in (D, E).

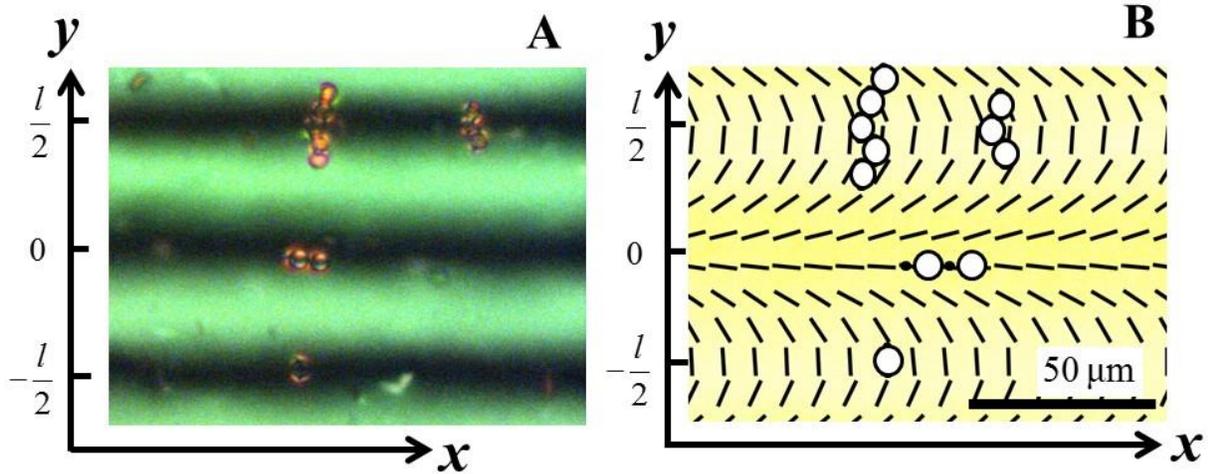

**Fig. 4. Preferential placement of colloids with different surface properties in a cell with periodically patterned splay-bend director field.** (**A**) Polarizing microscope texture of polystyrene colloids with homeotropic and tangential anchoring located in the alternating bands of splay and bend, respectively. (**B**) Scheme of preferential placement with the experimentally determined director pattern (PolScope observations). The homeotropically anchored spheres reside in the region of splay, $y = 0$, while tangentially anchored spheres localize in the regions of bend, $y = \pm l/2$. Note different orientation of chains formed by homeotropic spheres (along the $x$-axis) and by the tangential spheres (perpendicular to the $x$-axis). The colloids are placed in a LC mixture with zero dielectric anisotropy. P and A show the directions of polarizer and analyzer.